\documentclass{Interspeech}

\interspeechcameraready

\title{Comparative Evaluation of Acoustic Feature Extraction Tools for Clinical Speech Analysis}

\author[affiliation={1}]{Anna Seo Gyeong}{Choi}
\author[affiliation={2}]{Alexander}{Richardson}
\author[affiliation={2}]{Ryan}{Partlan}
\author[affiliation={2}]{Sunny X.}{Tang$^*$}
\author[affiliation={3}]{Sunghye}{Cho$^*$}

\affiliation{Information Science}{Cornell University}{USA}
\affiliation{Department of Psychiatry}{Northwell Health}{USA}
\affiliation{Linguistic Data Consortium, Department of Linguistics}{University of Pennsylvania}{USA}
\email{sc2359@cornell.edu, arichardson7@northwell.edu, rpartlan@northwell.edu, stang3@northwell.edu, csunghye@ldc.upenn.edu}


\usepackage{comment}

\begin{document}

\maketitle
{\renewcommand{\thefootnote}{}\footnotetext{* These authors contributed equally as senior authors.}}
\begin{abstract}

This study compares three acoustic feature extraction toolkits—OpenSMILE, Praat, and Librosa—applied to clinical speech data from individuals with schizophrenia spectrum disorders (SSD) and healthy controls (HC). By standardizing extraction parameters across the toolkits, we analyzed speech samples from 33 SSD and 38 HC participants and found significant toolkit-dependent variations. While F0 percentiles showed high cross-toolkit correlation (r=0.962–0.999), measures like F0 standard deviation and formant values often had poor, even negative, agreement. Additionally, correlation patterns differed between SSD and HC groups. Classification analysis identified F0 mean, HNR, and MFCC1 (AUC \textgreater 0.70) as promising discriminators. These findings underscore reproducibility concerns and advocate for standardized protocols, multi-toolkit cross-validation, and transparent reporting.

\end{abstract}
\vspace{-0.5em}
\section{Introduction}

Acoustic features play a central role in speech processing pipelines, underpinning tasks such as speaker recognition \cite{sambur1975selection, li2022acoustic}, emotion classification \cite{rong2009acoustic, atmaja2022survey}, and clinical speech assessment \cite{michaelis1998selection, mittal2021machine}. Although end-to-end neural architectures have been increasingly adopted, classical feature-based approaches are still widely used. In clinical contexts, for example, speech signals are often collected in limited quantities due to time, cost, and logistical constraints. Under these conditions, hand-crafted features -- whose meaning can be grounded in linguistic and phonetic theory -- still provide an interpretable and data-efficient alternative and are sometimes used in combination with more advanced architectures \cite{omeroglu2022multi, pan2020acoustic}.

Despite the continued usage of hand-crafted acoustic features, there is a growing concern that practitioners often use multiple feature sets without fully understanding or validating each feature’s definition, extraction parameters, or applicability. Moreover, the availability of numerous open-source toolkits further complicates the reproducibility and consistency of extracted features. Each toolkit has different underlying assumptions, default settings, and target domains, potentially leading to contradictory or misleading results if not carefully reconciled. In particular, inconsistent or misleading results can have a significant impact in the clinical field, as future studies rely on previous findings to develop screening or monitoring tools for patients. 

In this paper, we conducted a comparative analysis of three popular acoustic feature extraction toolkits -- OpenSMILE \cite{eyben2010opensmile}, Praat \cite{boersma2001speak} (via its Python wrapper Parselmouth \cite{jadoul2018introducing}), and Librosa \cite{mcfee2015librosa} -- applied to a clinical speech dataset of people with Schizophrenia Spectrum Disorders (SSD) and healthy controls (HC). We aligned the parameters as closely as possible across all three tools. This paper aims to answer (1) if acoustic features extracted with different toolkits show consistent results, (2) which acoustic features are most robust across toolkits and participant groups, and (3) what toolkits are reliable to use in clinical studies. 
\vspace{-0.5em}
\section{Previous Studies}

The comparison and validation of acoustic feature extraction tools have been ongoing concerns in speech processing research \cite{shrawankar2013techniques, kurzekar2014comparative}, yet systematic evaluations specifically targeting clinical applications remain limited. Early comparative studies across different toolkits focused primarily on general speech analysis \cite{lenain2020surfboard, ozseven2018speech}, leaving a significant gap in understanding how these tools perform in clinical contexts. 

The challenge of reproducibility in clinical speech research has become increasingly apparent, with several studies reporting that different research groups using different tools often produce conflicting results even when analyzing similar populations \cite{stegmann2020repeatability, almaghrabi2022reproducibility}. This lack of consistency has raised concerns about the reliability of acoustic features as clinical markers and highlighted the need for standardization in feature extraction methodologies.

Recent work has also begun to explore how deep learning approaches might complement or replace traditional acoustic feature extraction \cite{best2023deep, hong2020combining}. However, the interpretability and theoretical grounding of classical acoustic features continue to make them valuable in clinical contexts, particularly when working with limited data or when explanatory insight is required. This ongoing relevance of traditional acoustic features makes the investigation of their reliability even more critical.

Despite various investigations, there remains a significant gap in our understanding of how different feature extraction tools perform in clinical speech analysis. The present study addresses this gap by providing a systematic comparison of three widely-used toolkits, with a specific focus on features relevant to clinical assessment and the particular challenges posed by pathological speech.
\vspace{-0.5em}
\section{Methods}
\vspace{-0.5em}
\subsection{Data collection}
Participants diagnosed with schizophrenia spectrum disorders (SSD; N = 33, females = 24.7\%, mean age = 35.92) were enrolled from both inpatient and outpatient departments at a hospital in the US. Participants underwent screening using the psychosis and mood modules of the Structured Clinical Interview for DSM-IV \cite{first2004structured} and were confirmed to meet DSM-5 diagnostic criteria for schizophrenia spectrum disorders. Healthy volunteers (N = 38, females = 51.7\%, mean age = 36.12) were also enrolled either through their previous involvement in other research studies or by responding to online advertisements. All participants gave informed consent, with minors providing assent. All study procedures were approved by the Institutional Review Board. All participants performed several speech-based tasks, including three picture description tasks, as part of a larger study. All speech samples were digitally recorded, and later manually time-stamped and annotated by trained human annotators. 
\vspace{-0.5em}
\subsection{Extraction tools}
We compared three widely-used acoustic feature extraction toolkits: OpenSMILE, Praat, and Librosa. Each has a distinct implementation approach:

\textbf{OpenSMILE} \cite{eyben2010opensmile} is designed for batch extraction of large feature sets for machine learning applications. OpenSMILE has been extensively used in various tasks such as speech emotion recognition \cite{mustafa2024automatic} and clinical speech analysis \cite{stegmann2020repeatability, lin2020identification}, as well as paralinguistic challenges \cite{ashok2022paralinguistic}. We used the eGeMAPS configuration \cite{eyben2015geneva} for the extraction, which has become a standard in affective computing and clinical speech research \cite{barche2020towards}.

\textbf{Praat} \cite{boersma2001speak} is primarily designed for interactive phonetic analysis. It employs algorithm-specific implementations for each feature type, with a focus on accuracy over computational efficiency. Praat remains the gold standard in clinical phonetics \cite{stegmann2020repeatability, maryn2017practical} and detailed acoustic-phonetic studies in the broader Linguistics field \cite{styler2013using}, particularly when precise voice quality measurements are needed.

\textbf{Librosa} \cite{mcfee2015librosa} is a Python package originally designed for music information retrieval but increasingly used for speech processing. It employs probabilistic approaches for most of its features and emphasizes perceptual relevance, with its default settings often differing from speech-specific tools. Recently, Librosa has gained popularity in speech analysis for deep learning applications \cite{bhavya2022deep} and has also been used in clinical speech assessment \cite{huang2023evaluating, banks2024clinical} due to its integration with Python-based machine learning frameworks.

The three toolkits were clearly built and optimized for different purposes -- Praat for phonetic analysis, OpenSMILE for machine learning applications, and Librosa for music information retrieval. However, all of them have been frequently used in the literature, potentially contributing to mixed results.
\vspace{-0.5em}
\subsection{Feature Extraction Configuration}
To ensure fair comparison, we standardized the extraction parameters across all toolkits:
\begin{itemize}
    \item Sampling rate: 16kHz across all toolkits
    \item Frame size: 60ms (equivalent to 960 samples at 16kHz)
    \item Hop size: 10ms (160 samples)
    \item Window function: Hamming window
    \item Pre-emphasis: Disabled for consistency
    \item Frequency range: 0-8000Hz (Nyquist frequency)
    \item F0 search range: 55-1000Hz
    \item Silence thresholds: -60dB for voice activity detection
\end{itemize}
All other feature-specific thresholds were matched where possible. Despite our efforts to employ consistent parameters, certain toolkit-specific differences remained unavoidable, such as different underlying algorithms for F0 extraction (cross-correlation-based in OpenSMILE and Praat versus probabilistic in Librosa). All recordings were processed using the standardized pipeline described above, with identical parameter configurations applied consistently across all toolkits. The feature extraction process generated multiple statistics for each acoustic parameter (means, standard deviations, percentiles), which were then used in our correlation analysis.
\vspace{-0.5em}
\subsection{Acoustic Features and their Clinical Relevance}
Acoustic features provide an objective means to quantify speech patterns that may be altered in clinical populations. In this section, we describe the key features extracted in our study and discussed their relevance to clinical speech assessment, particularly in the context of schizophrenia.

\textbf{Fundamental frequency (F0)} represents the lowest frequency of vocal fold vibration during phonation. F0 characteristics are crucial in clinical linguistics as they reflect both physiological aspects of voice production and prosodic patterns that may be altered in various conditions. In schizophrenia, research has demonstrated abnormal prosodic patterns, often marked by reduced pitch variability and monotonous speech \cite{parola2023voice, compton2018aprosody}.

\textbf{Formants (F1-F3)} are local maxima in the spectrum that result from the acoustic resonance of the human vocal tract. Their values mostly correspond to the position of the tongue and other articulators during vowel production. In clinical populations, alterations in formant patterns may reflect abnormalities in articulation precision or stability. Research has shown that individuals with schizophrenia may exhibit shifts in vowel space areas and less stable formant trajectories, potentially due to motor control deficits or cognitive factors affecting speech planning \cite{hogoboom2023initial}. The relationship between formants -- particularly the ratios and distances between F1, F2, and F3 -- provides information about overall vocal tract configuration and motor planning and may serve as biomarkers for certain speech disorders \cite{shellikeri2024digital}.

\textbf{Harmonics-to-Noise Ratio (HNR)} quantifies the relative amount of periodic (harmonic) versus aperiodic (noise) components in the voice signal. This measure is particularly relevant for assessing voice quality, with lower values indicating increased breathiness, roughness, or general dysphonia. In schizophrenia research, reduced vocal quality -- possibly related to medication effects, smoking prevalence, or neuromotor factors -- has been documented using HNR measures \cite{zhao2022vocal}. 

\textbf{Jitter} and \textbf{shimmer} represent cycle-to-cycle variations in frequency and amplitude, respectively. These perturbation measures are highly sensitive to neuromotor control of the vocal folds and have been widely used in clinical voice assessment \cite{teixeira2016algorithm}. In schizophrenia, elevated jitter and shimmer values may indicate reduced precision in laryngeal control \cite{newman1989vocal}, potentially related to the broader motor coordination issues associated with the disorder. 

\textbf{Amplitude} captures the perceived intensity of speech, reflecting both physiological factors (respiratory support, vocal effort) and communicative intent (emphasis, emotional expression). In clinical populations, abnormal amplitude patterns -- whether monotonous speech with minimal intensity variation or inappropriate amplitude modulation -- can significantly impact communicative effectiveness. Individuals with schizophrenia often exhibit reduced amplitude variation consistent with negative symptoms like flat affect \cite{compton2018aprosody}.

\textbf{Mel-frequency cepstral coefficients (MFCCs)} provide a compact representation of the speech spectrum that roughly corresponds to the auditory system's response. The first MFCC (MFCC1) primarily reflects overall spectral energy distribution, while higher coefficients (MFCC2-4) capture increasingly fine spectral details related to vocal tract configuration and articulation. In schizophrenia research, altered MFCC patterns have been associated with both production differences (potentially related to articulatory precision or vocal tract tension) and perceptual characteristics that listeners identify as ``disorganized'' speech \cite{chakraborty2018prediction, zhang2016clinical}.

The clinical utility of these acoustic features extends beyond simple group discrimination to monitoring treatment effects, predicting functional outcomes, and potentially serving as objective biomarkers. However, this potential can only be realized if the features can be extracted reliably and consistently across different software implementations -- the central focus of our present investigation.
\vspace{-0.5em}
\subsection{Statistical analysis}
The analysis involved several statistical approaches to evaluate consistency and reliability across the three toolkits. Pearson correlation coefficients were calculated between each toolkit pair for every acoustic feature to assess agreement. Statistical significance of correlations was determined using standard t-tests with p-value thresholds of 0.05, 0.01, and 0.001. To evaluate whether correlations differed significantly between toolkit pairs, Fisher's r-to-z transformation was applied, comparing correlation coefficients with rigorous statistical testing. Additionally, the classification potential of features for distinguishing SSD from HC groups was assessed using ROC curve analysis and calculating Area Under Curve (AUC) values, with features showing AUC values above 0.7 considered to have good discrimination potential.

\vspace{-0.5em}
\section{Results}

\begin{figure*}[!ht]
\begin{center}
\includegraphics[width=\textwidth]{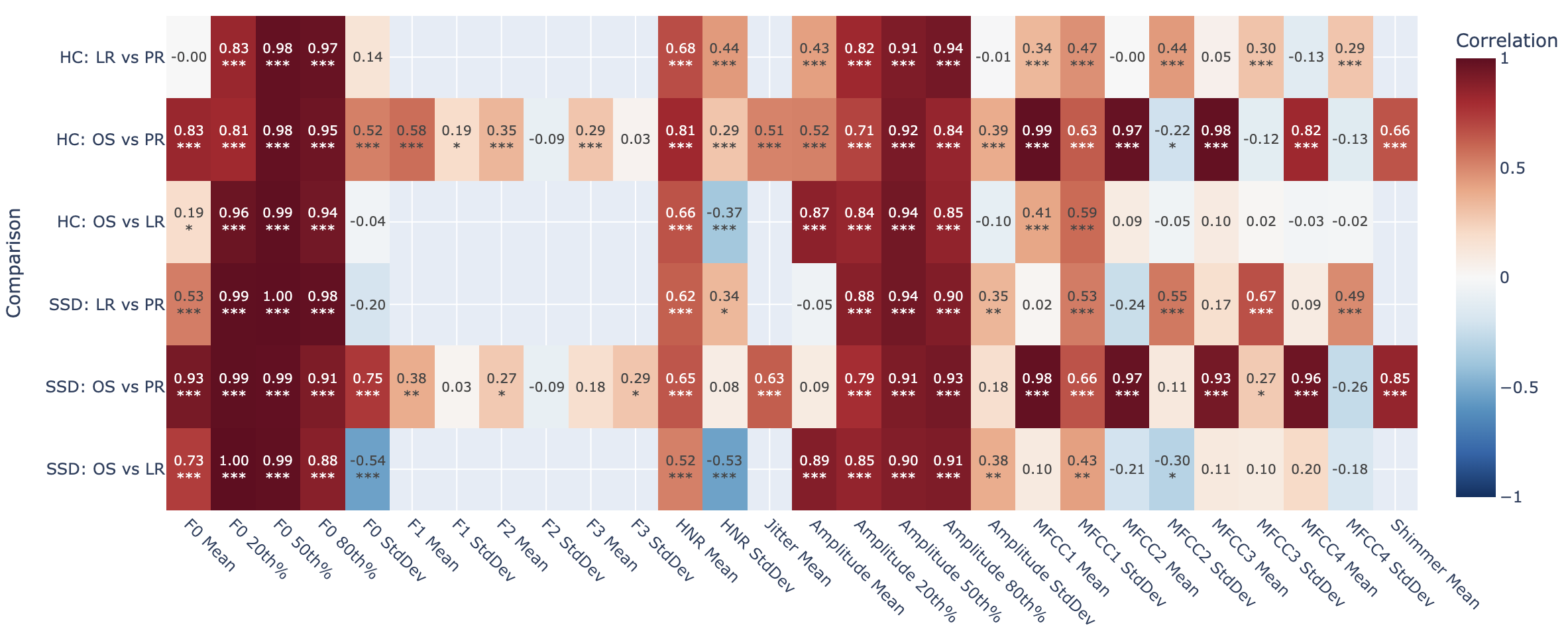}
\caption{Correlation heatmap comparing acoustic feature extraction across three tools (OpenSMILE [OS], Praat [PR], and Librosa [LR]) for SSD and HC groups. Color intensity indicates correlation strength from -1 (dark blue) to 1 (dark red). Statistical significance of correlations is marked: * (p\textless0.05), ** (p\textless0.01), and *** (p\textless0.001). Empty cells indicate toolkit pairs not available in specific features.}
\label{corr_heatmap}
\end{center}
\end{figure*}

The correlation analysis reveals distinct patterns of agreement between the three feature extraction tools. Figure \ref{corr_heatmap} presents the correlation matrix for both the SSD and HC groups after removing outliers at the 25th and 75th percentiles.

For F0 percentile measurements, we observe remarkably high correlations, particularly between OpenSMILE and Librosa (r=0.993-0.999 for SSD group, p\textless0.001; r=0.962-0.993 for HC group, p\textless0.001). The correlation between OpenSMILE and Praat is similarly strong (r=0.988-0.994 for SSD, p\textless0.001; r=0.809-0.981 for HC, p\textless0.001). However, F0 mean shows more moderate correlation between OpenSMILE and Librosa (r=0.730 for SSD, p\textless0.001; r=0.189 for HC, p\textgreater0.05), suggesting algorithm-specific differences in handling of unvoiced frames or edge conditions. Most notably, F0 standard deviation exhibits poor correlation between tools, with negative correlations between OpenSMILE and Librosa (r=-0.536 for SSD, p\textless0.001; r=-0.040 for HC, p\textgreater0.05) and between Librosa and Praat (r=-0.197 for SSD, p\textgreater0.05; r=0.144 for HC, p\textgreater0.05). This discrepancy likely stems from fundamental differences in the underlying F0 extraction algorithms and how they handle voice onset/offset transitions.

Formant measurements (F1-F3) show inconsistent extraction across toolkits, suggesting substantial differences in formant estimation algorithms across tools.

Harmonics-to-Noise Ratio (HNR) shows moderate correlation between OpenSMILE and Praat (r=0.649 for SSD, p\textless0.001; r=0.813 for HC, p\textless0.001) and between Librosa and Praat (r=0.622 for SSD, p\textless0.001; r=0.677 for HC, p\textless0.001). However, HNR standard deviation exhibits poor correlation between OpenSMILE and Librosa (r=-0.534 for SSD, p\textless0.001; r=-0.374 for HC, p\textless0.001) and only moderate correlation between OpenSMILE and Praat (r=0.084 for SSD, p\textgreater0.05; r=0.289 for HC, p\textless0.05).

Jitter measurements show reasonable correlation (r=0.629 for SSD, p\textless0.001; r=0.512 for HC, p\textless0.001) between OpenSMILE and Praat. Similarly, shimmer values exhibit good correlation (r=0.846 for SSD, p\textless0.001; r=0.658 for HC, p\textless0.001) between OpenSMILE and Praat.

Amplitude mean shows strong correlation between OpenSMILE and Librosa (r=0.892 for SSD, p\textless0.001; r=0.869 for HC, p\textless0.001) but poorer agreement with Praat (r=-0.052 for SSD vs Librosa, p\textgreater0.05; r=0.425 for HC vs Librosa, p\textless0.001). Amplitude percentiles demonstrate more consistent correlation across all three tools, particularly for the SSD group (r=0.847-0.948, p\textless0.001). Interestingly, amplitude standard deviation shows much weaker agreement, especially between OpenSMILE and Praat (r=0.177 for SSD, p\textgreater0.05; r=0.386 for HC, p\textless0.001).

The correlation patterns for MFCCs vary considerably by coefficient number. MFCC1 mean shows high correlation between OpenSMILE and Praat (r=0.981 for SSD, p\textless0.001; r=0.989 for HC, p\textless0.001) but low correlation between OpenSMILE and Librosa (r=0.100 for SSD, p\textgreater0.05; r=0.406 for HC, p\textless0.001). Higher-order MFCCs (2-4) show more variable correlation patterns, ranging from strong agreement (r=0.974 for MFCC2 OpenSMILE vs Praat in HC, p\textless0.001) to negative correlations (r=-0.240 for MFCC2 Librosa vs Praat in SSD, p\textless0.05).

Notably, the correlation patterns differ between the SSD and HC groups for several features, with these differences being statistically significant (p\textless0.05) in features such as F0 mean, HNR mean, and amplitude percentiles. Classification analysis revealed that F0 mean, HNR measures, and MFCC1 features had the highest discrimination potential between SSD and HC groups, with AUC values above 0.75, particularly when extracted using OpenSMILE.


\vspace{-0.5em}
\section{Discussion \& Conclusion}
Our findings reveal significant inconsistencies in feature extraction results, despite careful parameter alignment and standardized processing. These results have important implications for clinical speech analysis and highlight several critical considerations for future research. 

The observed discrepancies in feature values raise serious concerns about the reproducibility and reliability of acoustic analyses in clinical applications. While some features demonstrated high cross-toolkit correlation, others showed poor or even negative correlations. Such inconsistencies may explain contradictory or mixed findings across studies using different toolkits. Our results highlight the importance of reporting detailed methodological information, including toolkits and parameters employed, to allow the reproducibility of findings in future research.

Of particular importance is the impact these findings have on interpretability in clinical settings. Unlike consumer applications where end-performance may be the primary concern, clinical contexts demand transparency and interpretability. Medical professionals need to understand why a classification or assessment was made to inform treatment decisions and communicate with patients. When acoustic features are inconsistently extracted, derived conclusions become questionable, regardless of classification accuracy. The interpretability of acoustic features -- their grounding in physiological and linguistic theory -- represents one of their key advantages over black-box approaches. However, this advantage is severely compromised when feature extraction itself lacks consistency.

Our results should serve as a call to action for the speech processing community, particularly researchers working in clinical domains. The current practice of extracting hundreds of features and blindly applying dimensionality reduction techniques without proper understanding of the underlying linguistic and physiological basis can be problematic and potentially misleading. This approach may become even more concerning as the field moves toward acoustic embeddings derived from deep neural networks, which rarely offer clinical interpretability. While such techniques may achieve high performance on specific datasets, they may provide little insight into the underlying speech characteristics and may perpetuate or mask extraction inconsistencies.

We recommend that practitioners consider taking the following steps forward:
\begin{enumerate}
    \item Development of standardized extraction protocols specifically designed for clinical speech analysis, with validated parameters across different toolkits
    \item Increased transparency in research publications about extraction methods, tool versions, and parameter configurations to improve reproducibility
    \item Cross-validation of acoustic features using multiple extraction tools and datasets before drawing clinical conclusions
    \item Greater collaboration between speech technology experts, linguists, and clinical practitioners to ensure feature interpretation is grounded in both technical accuracy and clinical relevance
    \item Critical evaluation of newer embedding approaches in clinical contexts, with careful consideration of the trade-off between performance and interpretability
\end{enumerate}

The field stands at a critical juncture where computational advances must be balanced against clinical needs for transparency and theoretical grounding. As automated speech analysis continues to gain traction in healthcare applications, ensuring reliable, interpretable, and consistent measurements becomes imperative. This work contributes to that goal by highlighting current limitations and calling for increased transparency and standardization in acoustic feature extraction for clinical speech analysis.

\vspace{-1em}
\bibliographystyle{IEEEtran}
\bibliography{mybib}

\end{document}